\documentclass[aip, pof, preprint, longbibliography]{revtex4-1}

\usepackage{graphicx}
\usepackage{array} 
\usepackage{multirow}
\usepackage{amsmath}
\usepackage{color}
\usepackage{natbib}
\usepackage{rotating}
\usepackage{verbatim}
\usepackage[normalem]{ulem}

\begin{document}
\title{Flow speed has little impact on propulsive characteristics of oscillating foils}

\author{T. Van Buren}
\email[Email address for correspondence: ]{tburen@princeton.edu}
\affiliation{Mechanical and Aerospace Engineering, Princeton University, Princeton NJ 08544, USA}

\author{D. Floryan}
\affiliation{Mechanical and Aerospace Engineering, Princeton University, Princeton NJ 08544, USA}

\author{N. Wei}
\affiliation{Mechanical and Aerospace Engineering, Princeton University, Princeton NJ 08544, USA}

\author{A.J. Smits}
\affiliation{Mechanical and Aerospace Engineering, Princeton University, Princeton NJ 08544, USA}

\begin{abstract}
Experiments are reported on the performance of a pitching and heaving two-dimensional foil in a water channel in either continuous or intermittent motion.  We find that the thrust and power are independent of the mean freestream velocity for two-fold changes in the mean velocity (four-fold in the dynamic pressure), and for oscillations in the velocity up to 38\% of the mean, where the oscillations are intended to mimic those of freely swimming motions where the thrust varies during the flapping cycle.  We demonstrate that the correct velocity scale is not the flow velocity but the mean velocity of the trailing edge. We also find little or no impact of streamwise velocity change on the wake characteristics such as vortex organization, vortex strength, and time-averaged velocity profile development---the wake is both qualitatively and quantitatively unchanged. Our results suggest that constant velocity studies can be used to make robust conclusions about swimming performance without a need to explore the free-swimming condition.
\end{abstract}

\maketitle

\section{Introduction}\label{S1}
Many fishes laterally oscillate their fins in order to propel themselves and this fin motion generates an unsteady propulsive force, which in turn produces an unsteady swimming speed and acceleration \citep{walker1997, walker2004, tytell2007, lauder2008}.  To study the unsteady hydrodynamics of fish-like swimming, however, most experimentalists simplify the problem and laterally oscillate foils in a flow with a fixed freestream velocity. An obvious question to ask is whether the behavior of a foil in a flow of fixed velocity accurately represents the behavior of a foil that is free to move as it accelerates periodically.  

The few studies that have addressed the effects of the free-swimming condition on the forces and energetics of a propulsor indicate that the differences from constant-velocity swimming may actually be rather small. For instance, \citealt{wen2013} found that adding a periodic streamwise motion to a heaving flexible foil in an otherwise constant velocity flow had no impact on its average power consumption, though the range of added streamwise motions considered was quite small. Similar results have been found for the performance of fish and fish models. For example, \citealt{borazjani2010} found in their simulations that the Strouhal and Reynolds numbers for fixed and free-swimming carangiform and anguilliform swimmers at zero net-thrust were similar and that the efficiency and power coefficients for these two conditions were also in good agreement. Similarly, \citealt{bale2014} found that the fluctuating component of the swimming speed in knifefish and larval zebrafish contributed very little to the total power.  

Simulations by \citealt{hieber2008} and \citealt{zhou2012} showed that as an anguilliform swimmer increases speed from rest to a steady-state free-swimming velocity, the net-thrust only depends very weakly on the mean swimming velocity, while the side-force shows no impact. A force analysis by \citealt{curet2011} successfully predicted the free-swimming speed of a robotic knifefish by balancing the thrust force generated while tethered in still water and body drag force as it varied with speed.

Despite these observations, it is not widely recognized that the time-averaged performance of unsteady propulsors is independent of the flow velocity, and that tethered and free-swimming conditions can often be conflated.   For example, \citealt{carling1998} specifically states that ``[m]odels that assume a constant forward speed cannot be used to reach reliable conclusions about the development of forces during swimming." Though this work considers anguilliform swimmers, it highlights a prevailing attitude toward free-swimming within the community.  For a more recent examples, see \citealt{Das2017, Ryu2017} and \citealt{young2017}.

Here we attempt to bring some clarity to this question by analyzing the impact of substantial changes in the freestream velocity on the mean propulsive performance characteristics and wake structure of pitching and/or heaving rigid foils. We consider foils oscillating in either continuous or intermittent motion, and we examine the effects of changing the mean velocity, as well as adding velocity oscillations by moving the foil in the streamwise direction sinusoidally with an amplitude of up to 38\% of the mean velocity, considerably higher than that seen in biology \citep{xiong2014}.  Although we study simple rigid propulsors in isolation, we expect our results to apply more broadly to aquatic swimmers where the main source of drag can be separated from the main source of thrust, such as thunniform or carangiform swimmers where the body (drag source) and the caudal fin (thrust source) can be distinctly identified.  Examples include tuna, mackerel, dolphin, and trout.

\section{Experimental setup}\label{S2}
\begin{figure}
\begin{center}
\includegraphics[width=1\textwidth]{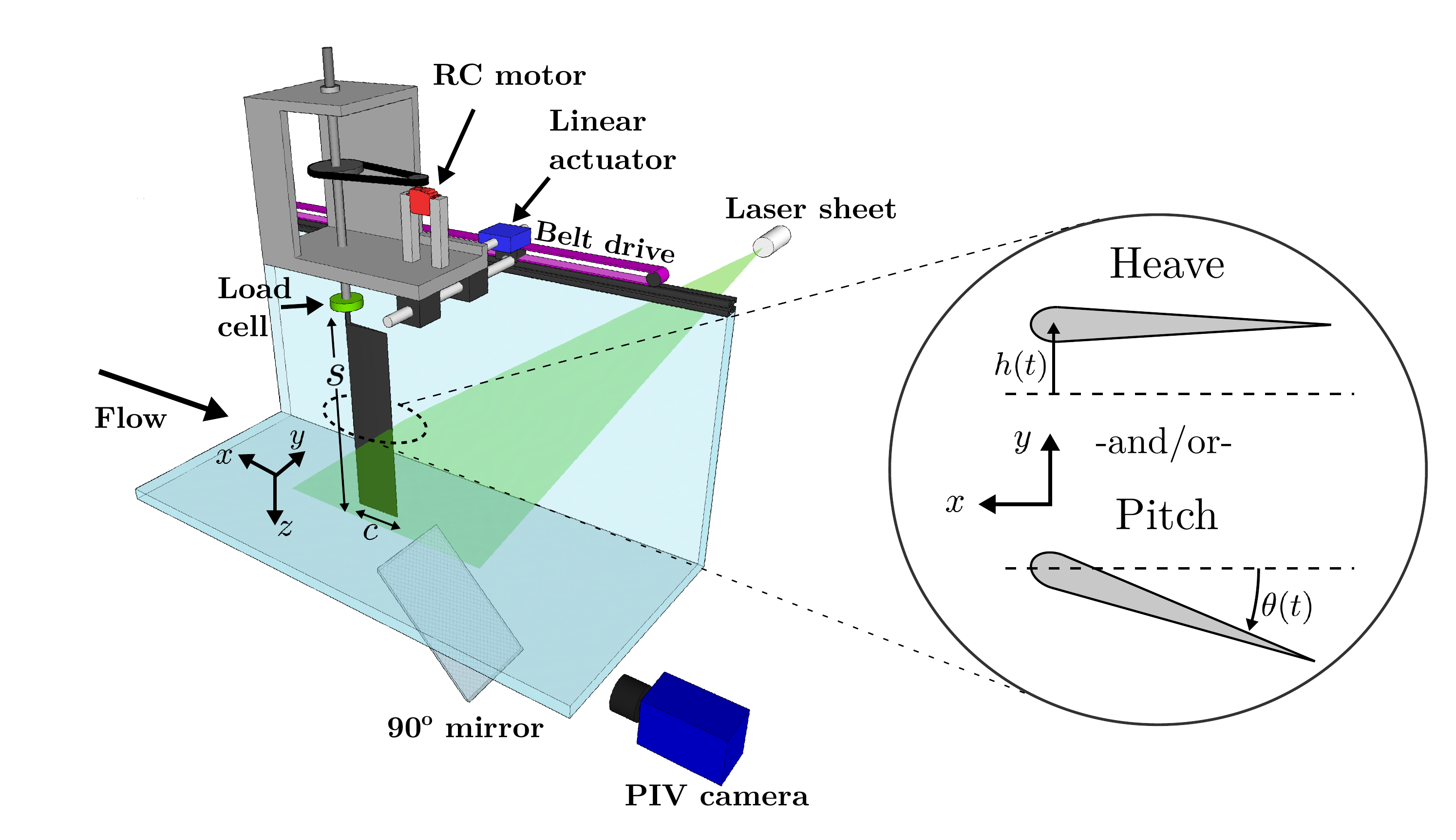}
\end{center}
\caption{Experimental setup.}
\label{expsetup}
\end{figure}

Experiments were conducted in a water channel on a pitching and/or heaving foil, as shown in figure~\ref{expsetup}. The water channel was a free-surface recirculating facility with a test section 0.46 m wide, 0.3 m deep, and 2.44 m long. Surface waves were minimized through baffles on the top surface, and the maximum turbulence intensity was 0.8\%. The mean freestream velocity was varied from $\overline{U}_\infty=60$ mm/s to 120 mm/s. A belt drive (Baldor BSM50N-375AF motor) was used to impose sinusoidal velocity oscillations by moving the foil in the streamwise direction with amplitude $u_a$, where $u_a/\overline{U}_\infty$ varied from 0 to 0.38, corresponding to maximum streamwise position changes of 40 mm (50\% of the chord). The resolution of streamwise movement is 0.292 mm per degree of motor rotation.

The propulsor was a two-dimensional teardrop foil with a chord length $c=80$ mm, maximum thickness 8 mm, and span $s=279$ mm. The foil was pitched about the leading edge via a RC-motor (Hitec HS-8370TH) and heaved with a linear actuator (Linmot PS01-23x80F-HP-R), monitored via encoders. Performance measurements consisted of: (1) pitch only, amplitudes $\theta_0=5^\circ$ to 15$^\circ$ in intervals of 1$^\circ$; (2) heave only, amplitudes $h_0/c=0.125$ to 0.25 in intervals of 0.0125; and (3) combined pitch heave motions, amplitudes $\theta_0=10^\circ$ and $h_0/c=0.0625$ to 0.1875 in intervals of 0.0125, where the pitching motion lagged the heaving motion by $90^\circ$. All imposed motions were sinusoidal. We consider both continuous and intermittent (duty cycle 0.5) swimming motions at fixed actuation frequencies of $f=0.75$ Hz and 1 Hz, respectively. All six components of forces and moments were monitored via a load cell (ATI Mini40) with force and torque resolutions of $5 \times 10^{-3}$ N and $1.25 \times 10^{-4}$ N$\cdot$m in the $x$- and $y$-directions respectively, and $10^{-3}$ N and $1.25 \times 10^{-4}$ N$\cdot$m in the $z$-direction, sampled at 1 kHz. Each time-averaged quantity consisted of an average of three separate trials of 20 actuation periods. 

As shown by the analysis presented in Appendix B, the inertia of the foil will have no impact on our results; the effects of inertia on the mean forces and power are exactly zero for the types of motion studied here.

Wake measurements were taken with two-component particle image velocimetry (PIV) on a measurement plane at the half-span of the foil. Silver coated hollow ceramic spheres (Potter Industries Inc. Conduct-O-Fil AGSL150 TRD) were used to seed the flow, illuminated using a 3500 mW gallium-nitride continuous laser (S3 Arctic Series). Images were captured and processed via a LaVision PIV system with a 5.5 mega-pixel sCMOS camera acquired at 50 Hz. Ten actuation periods were sampled for phase-averaging. Images were processed sequentially using a final spatial correlation interrogation window size of 64$\times$64 pixels with 50\% overlap. The final trimmed vector field grid size is 68$\times$80 velocity vectors. Average and instantaneous velocity errors were estimated to be 2.7\% and 1 to 5\%, respectively \citep{Sciacchitanoetal2013}.

\section{Performance Results}

\label{S3S1}
Here, we will present the experimental results on thrust and power in dimensional form, and the appropriate scaling will be discussed in the following section.  We first consider the performance of a pitching and/or heaving foil that is actuated sinusoidally and continuously, that is, with a duty cycle of one. 

\begin{figure}
\begin{center}
\includegraphics[width=1\textwidth]{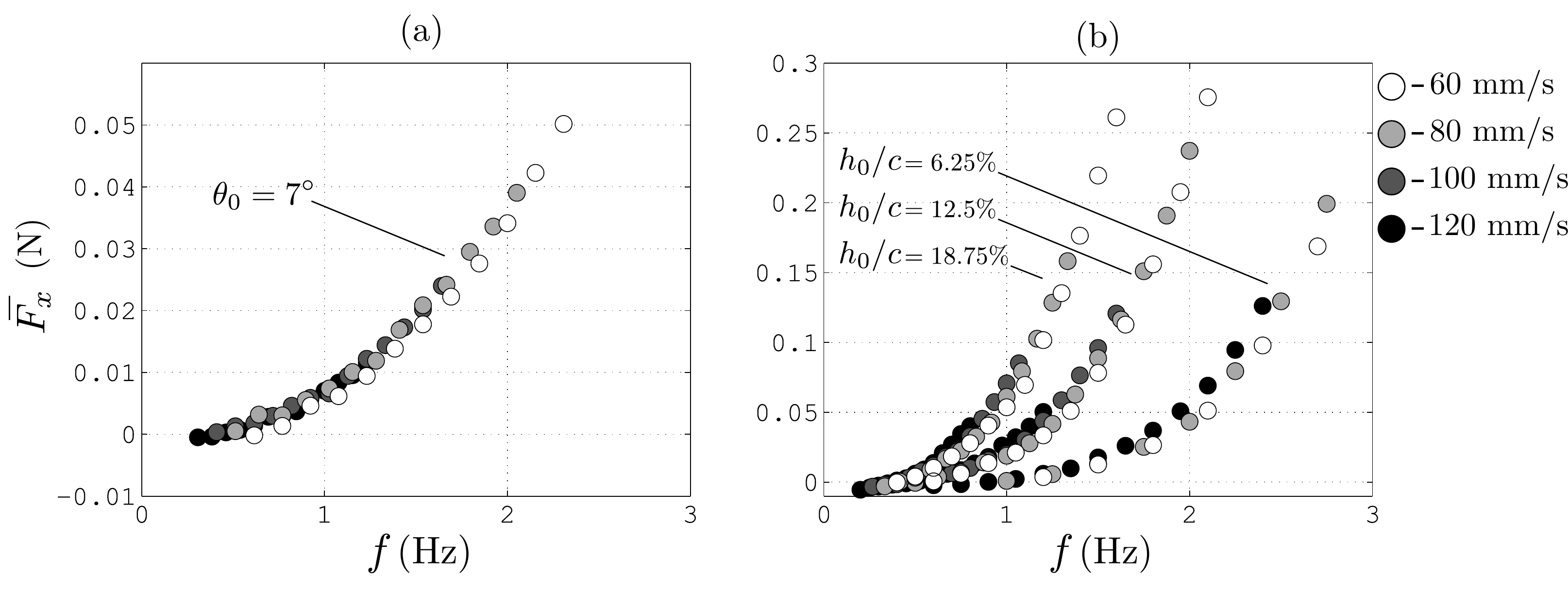}
\end{center}
\caption{Propulsor mean thrust force for continuous (a) pitching and (b) heaving motions for $\overline{U}_\infty=$ 60 mm/s (white circles); 80 mm/s (light grey circles); 100 mm/s (dark grey circles);  120 mm/s (black circles). One pitch and three heave amplitudes are shown. Adapted from \citealt{Floryan2016} with permission.}
\label{fig:meanU}
\end{figure}

The impact of changing the mean velocity $\overline{U}_\infty$ on the mean thrust produced $\overline{F}_x$ is shown in Figure \ref{fig:meanU}. For these cases, the foil was fixed in place and not allowed to move in the streamwise direction, and the mean velocity was varied over the range 60 mm/s to 120 mm/s.  In pitching motions, we see that the mean velocity has no measurable impact on the magnitude of the thrust. In heaving motions, there is little impact of mean velocity compared to relatively similar changes in foil kinematics (amplitude and frequency).

\begin{figure}
\begin{center}
\includegraphics[width=0.9\textwidth]{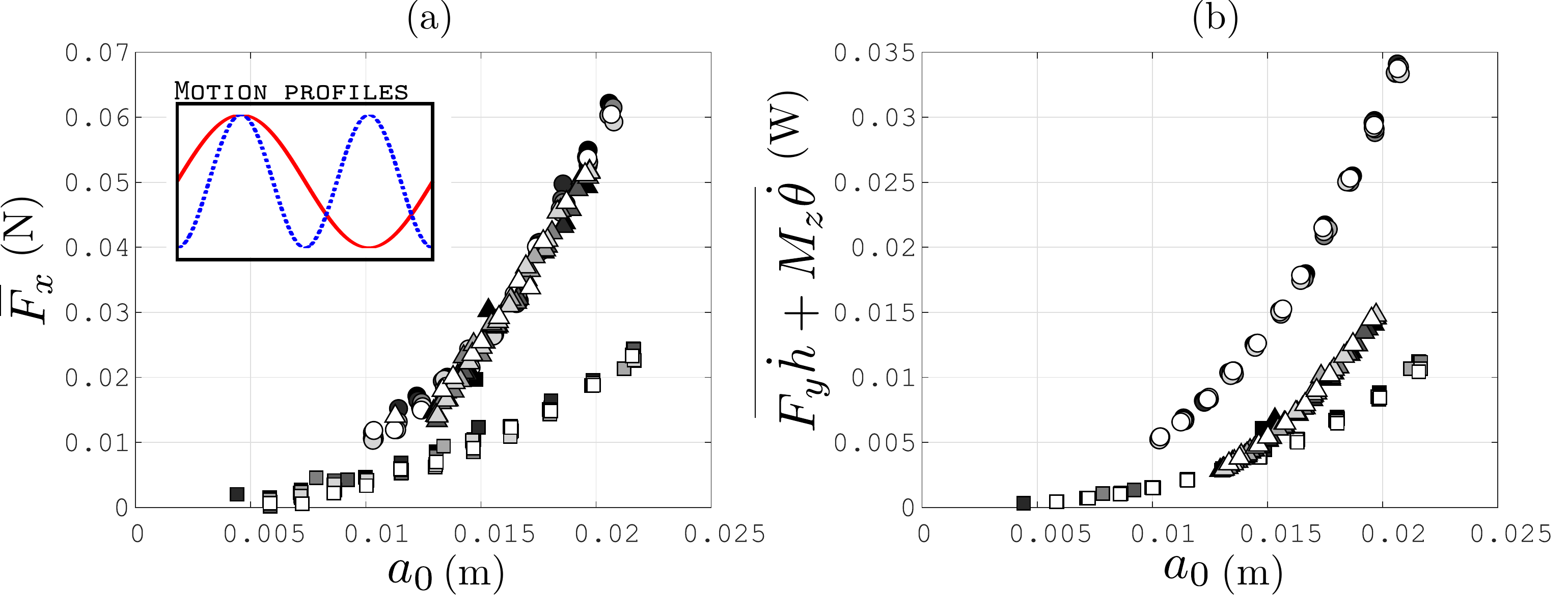}
\end{center}
\caption{Propulsor mean (a) thrust and (b) power undergoing streamwise velocity oscillations $0 \le u_a/\overline{U}_\infty \le 0.38$ (dark to light symbols), for $\overline{U}_\infty=80$ mm/s. Continuously heaving (circular symbols), pitching (square symbols), and pitching combined with heaving (triangular symbols) foil. Inlaid graphic represents the relative leading edge lateral position\textemdash or pitch angle for pitch only cases (red solid line), and streamwise position (dashed blue line).}
\label{fig:ContPerf}
\end{figure}

The effects of adding oscillations to the streamwise velocity while continuously pitching and/or heaving are shown in Figure \ref{fig:ContPerf} for $\overline{U}_\infty=80$ mm/s.  During continuous actuation, the thrust produced by the foil occurs at a frequency that is twice the actuation frequency due to the symmetry of the motion, so we imposed a streamwise velocity oscillation on the foil with a frequency that was twice the actuation frequency to mimic a naturally accelerating propulsor. Figure \ref{fig:ContPerf} shows the mean thrust and power for pitching and heaving motions. For velocity oscillation amplitudes up to 38\% of the freestream velocity, there is no discernible difference in performance from the case with no streamwise motion. 

\begin{figure}
\begin{center}
\includegraphics[width=0.9\textwidth]{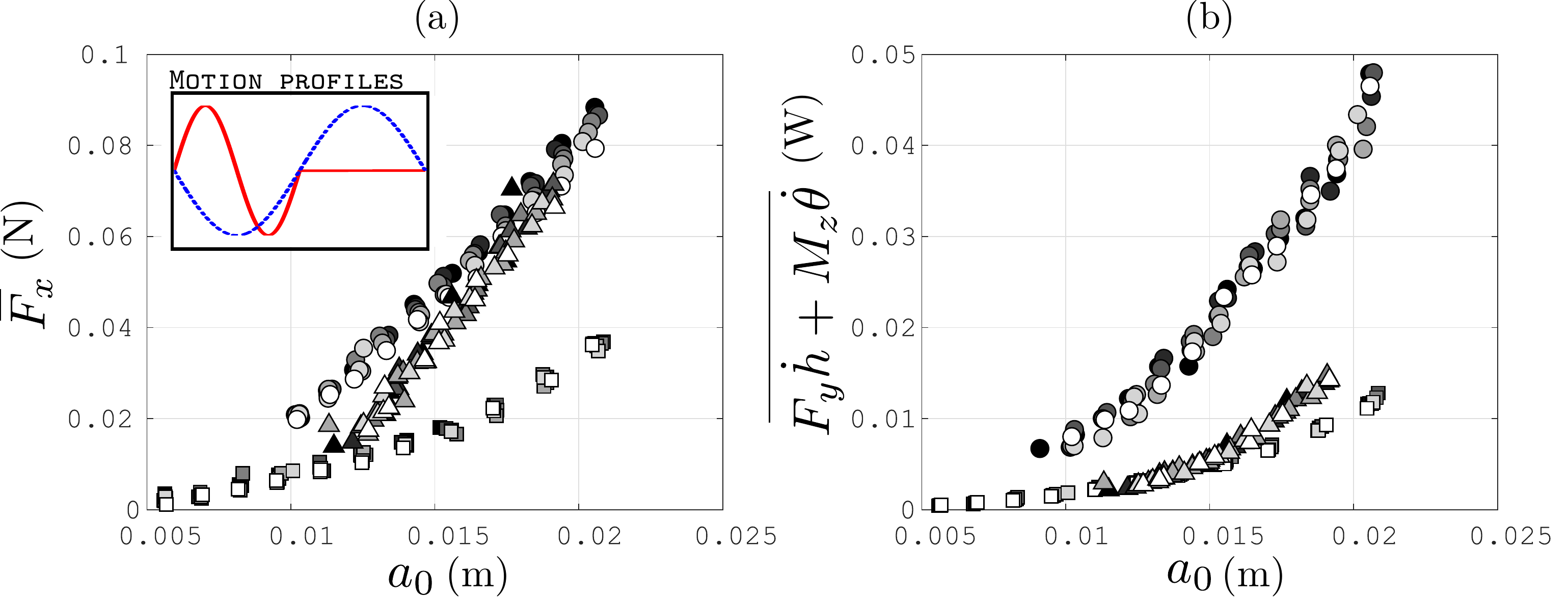}
\end{center}
\caption{Propulsor mean (a) thrust and (b) power undergoing streamwise velocity oscillations $0 \le u_a/\overline{U}_\infty \le 0.38$ (dark to light symbols). Intermittently heaving (circular symbols), pitching (square symbols), and pitching combined with heaving (triangular symbols) foil (duty cycle of 0.5). Inlaid graphic represents the relative leading edge lateral position\textemdash or pitch angle for pitch only cases (red solid line), and streamwise position (dashed blue line).}
\label{fig:BCPerf}
\end{figure}

We now consider the effects of intermittent actuation, that is, the effects of changing the duty cycle. For intermittent actuation, we imposed a streamwise velocity oscillation that mimicked burst and coast swimmers. The velocity oscillation frequency was set to half the actuation frequency so that the foil accelerated forward during the burst portion of the cycle and decelerated during the coast portion. For this experiment, we did not include the accelerations that would occur during the thrust cycle at twice the actuation frequency since the previous experiments demonstrated that these effects are negligible (see figure \ref{fig:ContPerf}). Figure \ref{fig:BCPerf} presents the thrust and power for intermittent pitching and/or heaving motions (duty cycle of 0.5), and we see that the performance of foils with intermittent actuation is not sensitive to substantial streamwise velocity oscillations. In all respects, the behavior is almost identical to that seen by \citealt{Floryan2017} in a study of a fixed foil undergoing intermittent actuation. 

\section{Scaling}

To understand these dimensional results better, we need to consider the appropriate non-dimensionalization, that is, the correct scaling parameters for the data. In most of the literature, the propulsive performance is represented by the thrust and power coefficients defined according to
\begin{equation}
C_T = \frac{F_x}{\frac{1}{2}\rho \overline{U}_\infty^2sc},\qquad C_P=\frac{F_y\dot{h}+M_z\dot{\theta}}{\frac{1}{2}\rho \overline{U}_\infty^3sc}
\label{perfEQ}
\end{equation}
where $F_x$ and $F_y$ are the streamwise and cross-stream forces acting on the foil, respectively (we call $F_x$ the thrust), $M_z$ is the spanwise moment, $\theta$ is the instantaneous pitch angle, $h$ is the instantaneous heave amplitude, and $\rho$ is the fluid density. The foil kinematics are characterized by the Strouhal number, $St=2fa/\overline{U}_\infty$, where $a$ is the peak amplitude of the trailing edge motion, and by the reduced frequency, $f^*=fc/\overline{U}_\infty$. 

These non-dimensional thrust and power coefficients are obviously not the correct parameters for the data presented here, in that any non-dimensionalization with respect to $\overline{U}_\infty$ would eliminate the collapse of the data in dimensional form.  Therefore we seek a more appropriate velocity scale.

In this respect, \citealt{Floryan2016} demonstrated that for pitching motions the thrust is due purely to added mass effects and the unsteady lift forces make no contribution.  Thus, the thrust for pitch is expected to scale as the component in the streamwise direction of the added mass ($\sim \rho c^2 s$) times the acceleration ($\sim c \ddot \theta$). That is, 
$$F_x \sim \rho s c^3 \ddot \theta \theta, $$ 
so that the mean thrust scales as 
\begin{equation}
 \overline{F}_x \sim \rho s c^3 f^2 \theta_0^2 \ \approx \rho s c f^2 a_0^2,
 \label{pitch_scale}
\end{equation}
where $a_0 \approx c\theta_0$ is the trailing edge amplitude for pitching motions. We see that the scaling suggests that the time-averaged thrust is independent of velocity, as borne out by the experimental results given in figure \ref{fig:meanU}a. 

Similarly, \citealt{Floryan2016} found that for heaving motions, the thrust is due purely to unsteady lift forces and that the added mass terms make no contribution to the thrust. Hence, the thrust for heave is expected to scale as the component in the streamwise direction of the instantaneous lift force. That is, 
$$F_x \sim L ( {\dot h}/{U^*} ) $$
where $L$ is the lift, $h$ is the instantaneous heave amplitude, and $U^*$ is the effective velocity seen by the foil. 
If we assume that the contribution to the lift is quasi-steady, and that the angle of attack $\alpha \approx {\dot h}/{U^*}$, for small $\alpha$ we obtain
\begin{equation*}
F_x   \sim {\textstyle \frac{1}{2}} \rho U^{*2} s c \, (2 \pi \alpha)  ( {\dot h}/{U^*} ) \sim  \pi \rho s c \, \dot h^2,
\end{equation*}
so that the mean thrust scales as 
\begin{equation}
\overline{F}_x \sim \rho s c f^2h_0^2 \ = \rho s c f^2a_0^2,
\label{heave_scale}
\end{equation}
where $a_0=h_0$ for heaving motions. This analysis indicates that the velocity does not appear in the leading order approximation of the mean thrust, and  the experimental results shown in figure \ref{fig:meanU}b demonstrate that this approximation holds well for the range of conditions studied here. The full unsteady analysis was given by \citealt{Floryan2016}.  The extension to simultaneous pitching and heaving motions was developed by \citealt{VanBuren2017_3}.

What about the effects of streamwise velocity perturbations? Let $g=b_0+b_1u'+b_2u'^2+...$ be the Taylor series expansion of the instantaneous thrust or power with respect to the perturbation velocity $u'$. When averaging in time, the first-order term integrates to zero since $u'$ is periodic. Thus, the effect of the perturbation velocity on the mean thrust or power is only at second-order and expected to be small for small values of $u'$. 

We see that the mean thrust forces developed by a pitching or heaving foil do not scale with dynamic pressure, in contrast to what might be assumed from aerodynamic considerations.  In fact, at the level of approximation adopted here, they do not depend on the mean velocity at all.  Instead, we find that the thrust for both pitching and heaving motions depends on the mean speed of the trailing edge, $V=fa_0$. We are not the first to suggest the importance of the lateral velocity scale.  The work by \citealt{garrick1936} on flapping and oscillating airfoils suggests that the mean thrust should depend approximately on $V^2$ (Garrick's equation 29 simplified), although this early result seems not to be widely known. In the context of fish swimming, \citealt{bainbridge1963} indicated that the thrust should depend on ``the square of its speed of transverse movement'', but the reasoning is unclear.  More recently, \citealt{gazzola2014} offered a mechanistic basis for the importance of the transverse tail velocity, and they showed that for added mass forces the thrust should scale as $V^2$. However, as \citealt{Floryan2016} demonstrated, aerodynamic forces are important when heaving motions are present, and so considerations of pitching and heaving propulsors need to take into account both added mass and lift-based forces. Our experiments substantiate the primacy of the lateral velocity scale over the streamwise velocity.

\begin{figure}
\begin{center}
\includegraphics[width=0.9\textwidth]{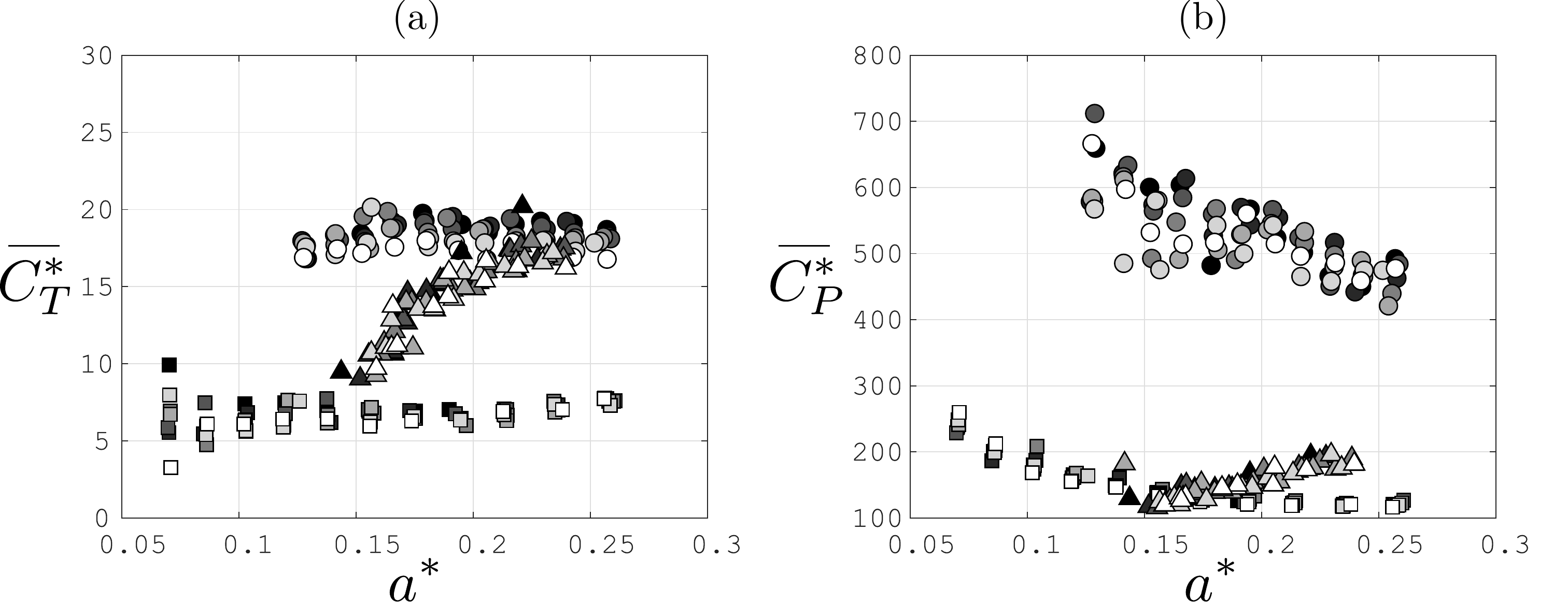}
\end{center}
\caption{Propulsor mean (a) thrust and (b) power non-dimensionalized by the trailing edge velocity. Symbols and colors as in figure \ref{fig:BCPerf}.}
\label{fig:BCPerfScaled}
\end{figure}

We therefore define new thrust and power coefficients according to
\begin{equation}
C_T^* = \frac{F_x}{\rho (f a_0)^2 s c}\qquad
C_P^* = \frac{P}{\rho (f a_0)^3 s c}.
\end{equation} 
As an example, figure \ref{fig:BCPerfScaled} shows the data of figure \ref{fig:BCPerf} using this scaling, where $a^*=a_0/c$. Note that the plots appear noisier because we are non-dimensionalizing by a measured output, $a_0$, and errors will go as $1/a_0^3$ and $1/a_0^4$ for the new thrust and power coefficients, respectively. This is more apparent at small $a_0$.

\section{Wake structure}\label{S3S2}

For free-swimming full-bodied fish, the wake changes with swimming speed \citep{tytell2004, drucker2000} because the propulsor thrust balances the drag of the body, which is a function of  swimming speed. Thus, different swimming speed requires different thrust, resulting in different tail kinematics and wake structure.  Here, instead, we consider an isolated foil, and we examine the impact of oscillations in the streamwise velocity on the wake without changing the foil kinematics.

\begin{figure}
\begin{center}
\includegraphics[width=0.9\textwidth]{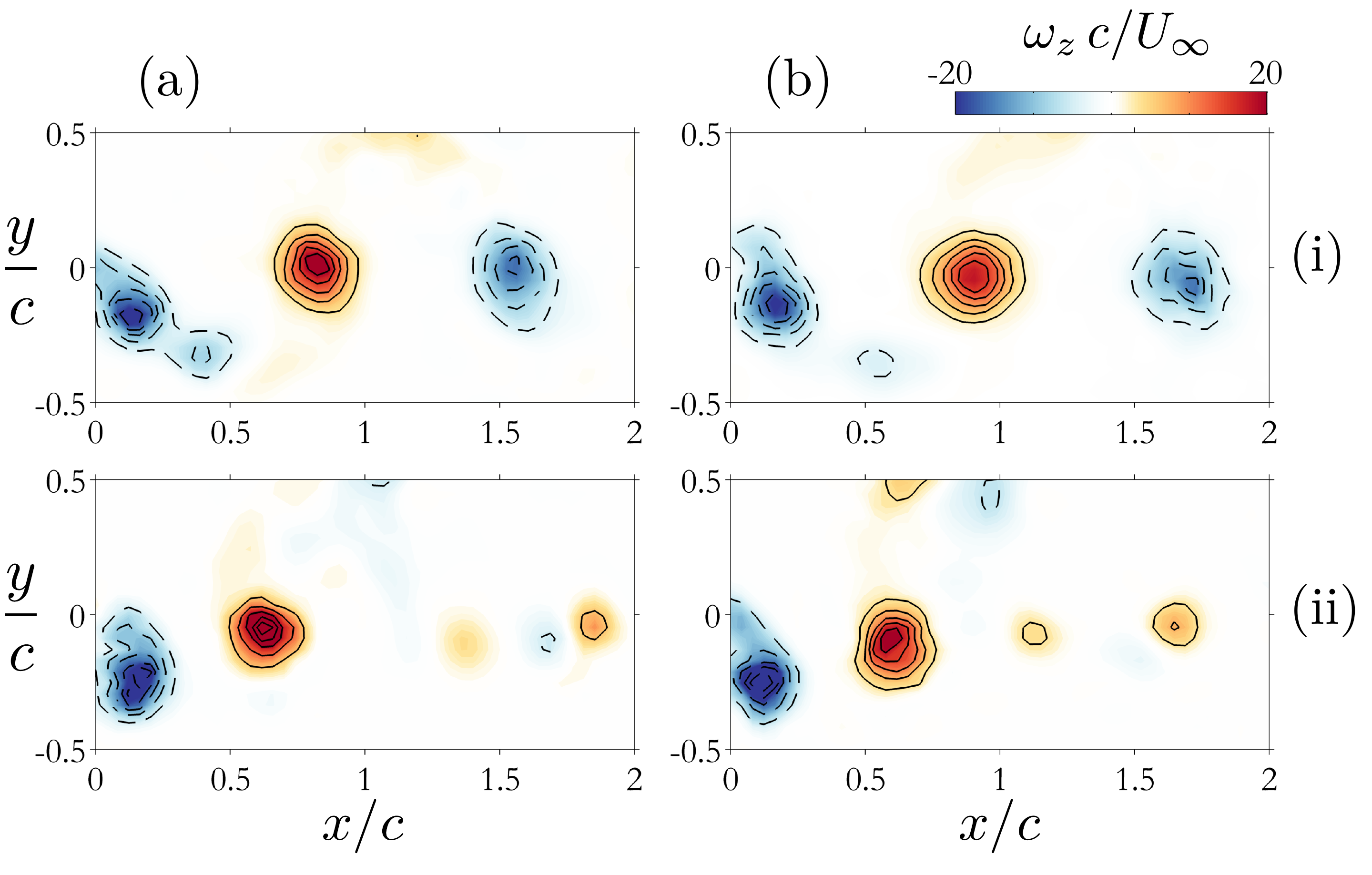}
\end{center}
\caption{Phase-averaged vorticity for (i) continuous and (ii) intermittent heaving motions at streamwise velocity amplitudes (a) $u_a/\overline{U}_\infty=0$ and (b) 0.38. Phase angle of heaving motion $\psi=0^\circ$ (passing through the neutral position moving into the positive $y$ domain).}
\label{fig:PIV}
\end{figure}

Figure \ref{fig:PIV} shows the phase-averaged vorticity distributions in the wake for the continuous and intermittent heaving motions with a steady freestream velocity ($u_a/\overline{U}_\infty=0$) and with the largest streamwise velocity oscillation explored here ($u_a/\overline{U}_\infty=0.38$). We see the expected reverse von K\'{a}rm\'{a}n vortex street for the continuous motions, and for the intermittent motions the wake is comprised of a vortex pair generated by the active portion of the cycle, surrounded by small secondary structures, consistent with the observations by \citealt{Floryan2017} of the wake of an intermittently actuated foil fixed in the streamwise direction. There is no discernible impact of the addition of an oscillating streamwise velocity on the shed vorticity field when comparing figure \ref{fig:PIV}i.a to i.b for continuous motions and figure \ref{fig:PIV}ii.a and ii.b for intermittent motions. A similar result was obtained for pitching motions.

These results are actually not surprising. Regardless of whether the freestream velocity is constant or changing in time, the vortices shed into the wake will be generated at twice the actuation frequency of the foil and at the same phase in the actuation cycle. The convection speed of the vortices will continue to be (approximately) equal to the mean freestream velocity, and as long as the mean freestream velocity does not change the spacing of the vortices is not expected to change either. Hence, the organization of the vortices should not be affected by the streamwise motion of the foil after the vortices are created; the vortices merely convect away from the foil at about the same speed as in the case where the freestream velocity is constant.

The strengths of the vortices are also unaffected. For continuous heaving, the total magnitude of circulation of two counter-rotating vortices at the phase angle of heaving motion $\psi=0^\circ$ (at $x/c=0.75$ and 1.6 in figure \ref{fig:PIV}.i) is $\Gamma=(\overline{U}_\infty c)^{-1}\iint_S |\omega_z| \cdot dS = 1.53$ and 1.50 for $u_a/\overline{U}_\infty=0$ and 0.38, respectively. Similarly, for continuous pitching (not shown for brevity), $\Gamma=1.38$ and 1.40 for the steady and most unsteady cases. These differences are within 2\%, which is the limit of our measurement accuracy.

The study by \citealt{fernando2016} on impulsively moving disks is helpful in explaining these observations. In their work, an elliptical plate was impulsively set into motion in a direction perpendicular to its surface; this is similar to our pitching and heaving foils moving in the cross-stream direction. They found that the circulation of the vortex created by the impulsive start initially grows as $\Gamma\sim d\,U_p$, where $d$ is the distance traveled and $U_p$ is the plate velocity. Relating this to an oscillating foil, the analog of $d$ is the amplitude of oscillation, $a$, and the analog of $U_p$ is the velocity of oscillation, $fa$. This suggests that the circulation of a vortex formed by an oscillating foil should scale as $\Gamma \sim fa^2$. Also, the area of a vortex should scale as $a^2$, in which case the strength of the vortex will scale as $f$. Thus, we would not expect the size and strength of the vortices created by the foil to be a function of the freestream velocity\textemdash only the amplitude and frequency of oscillation.

\begin{figure}
\begin{center}
\includegraphics[width=1\textwidth]{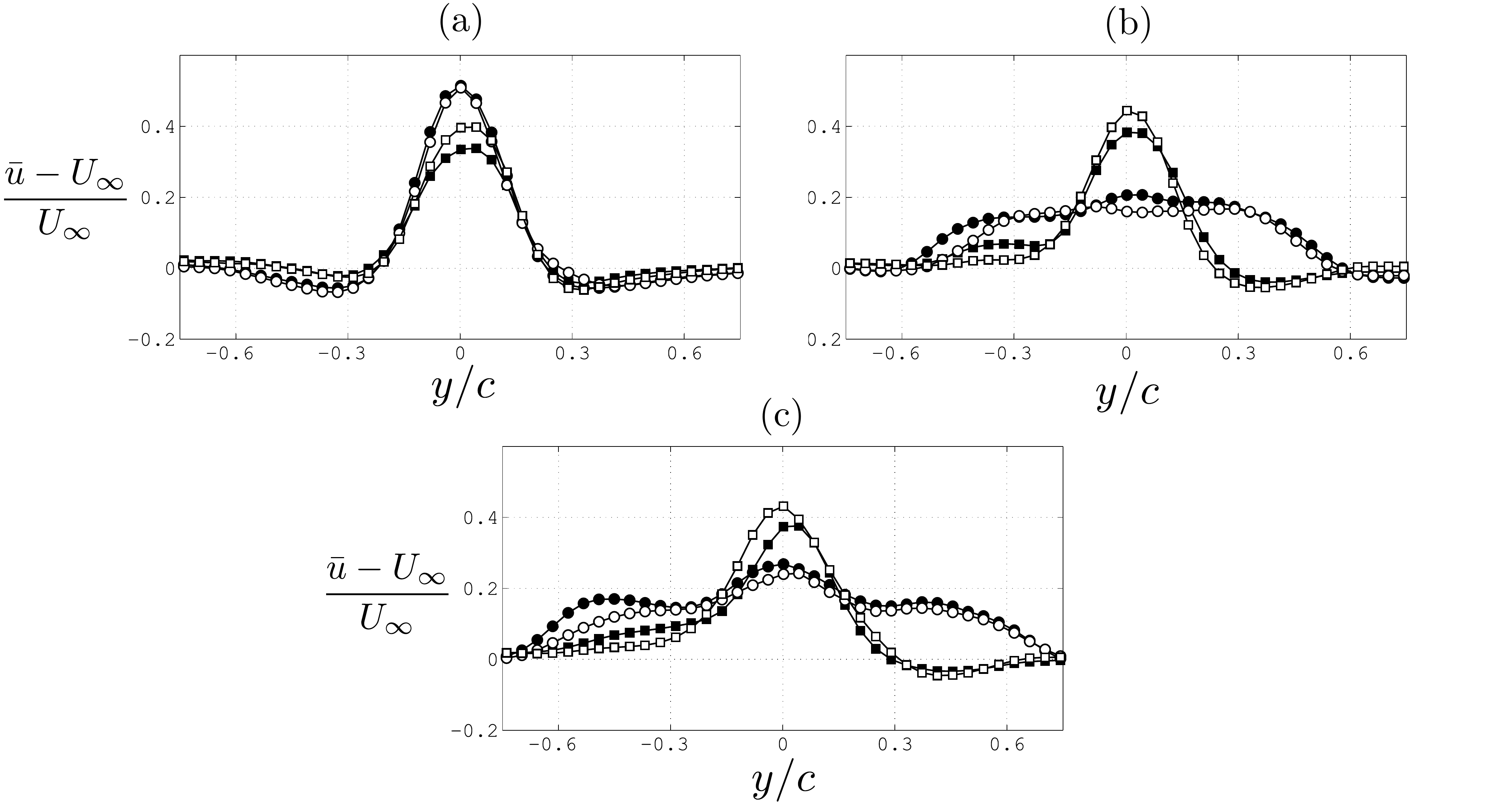}
\end{center}
\caption{Time-averaged streamwise velocity profiles at (a) $x/c=0.25$, (b) 1, and (c) 1.75 for continuous heaving (circular symbols) and pitching (square symbols) motions. Streamwise velocity amplitudes $u_a/\overline{U}_\infty=0$ (dark symbols) and 0.38 (light symbols).}
\label{fig:veloProf}
\end{figure} 

Given these results, we would expect the time-averaged velocity profiles to be also unaffected by the unsteady motion in the streamwise direction. As shown in figure \ref{fig:veloProf}, the velocity profiles indeed remain similar to one another throughout the downstream development, as the initial peak in added velocity diffuses and decays.

These wake results agree well with the performance characteristics presented in section \ref{S3S1}. The foil is producing the same strength vortices at the same spacing, which produces the same velocity profile and so the thrust and power produced by the foil remain unchanged.

\section{Applications to Biology}

Consider the dynamics of a freely swimming fish where the drag from the body and the thrust from the propulsor are distinct. The periodic forward swimming speed is given by
\begin{equation}
  m\frac{du}{dt} = T - D,
\end{equation}
where $m$ is the mass of the fish, $u$ is its forward swimming speed, $T$ is the thrust produced by its propulsor (for example, its caudal fin), and $D$ is the fluid drag experienced by the fish. As we derive in Appendix A, the mean speed is given by the balance of mean thrust and mean drag, \emph{minus} a modification due to the unsteady part of the thrust. This modification is generally small, thus, if we understand how the mean thrust is affected by the swimming speed (both the mean swimming speed and oscillations about the mean), it would seem that we may accurately predict the mean swimming speed  
for aquatic animals where the thrust-producing propulsor is distinct from the drag-producing body, such as thunniform and carangiform swimmers.  

\begin{figure}
\begin{center}
\includegraphics[width=1\textwidth]{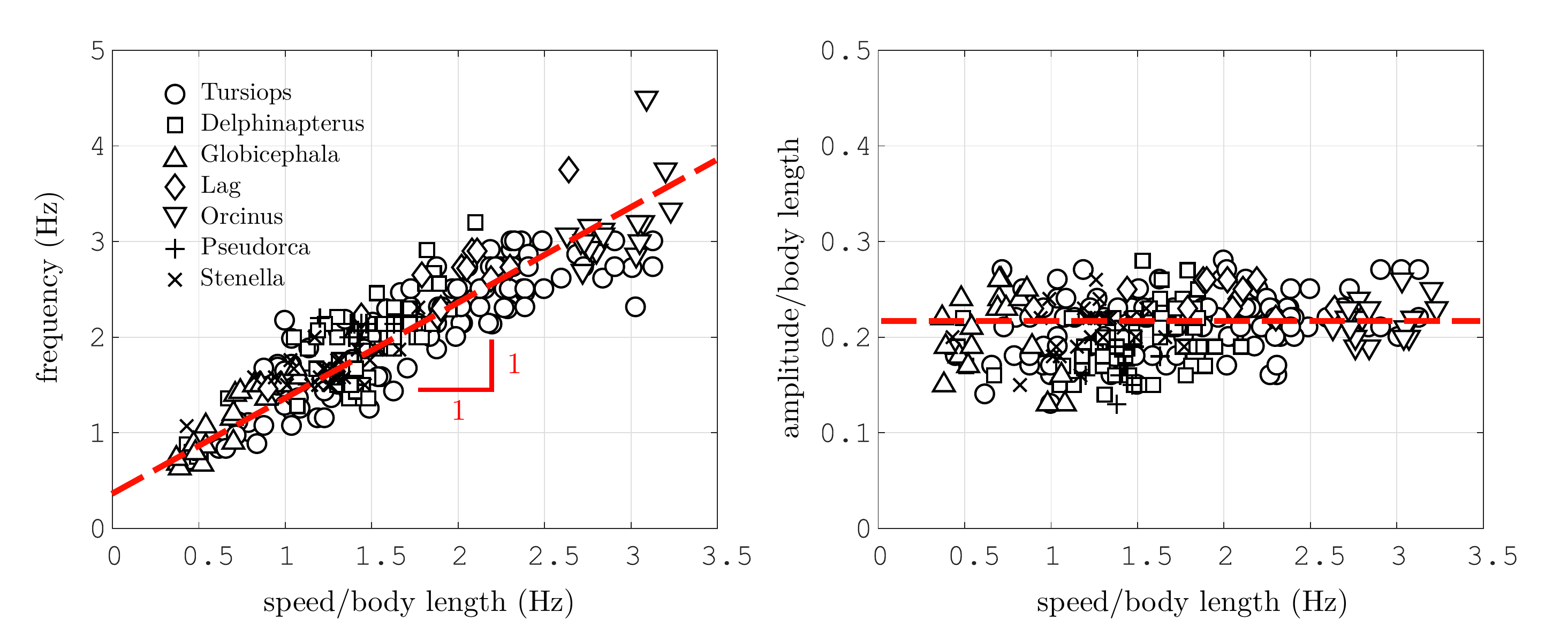}
\end{center}
\caption{ (a) Fluke-beat frequency and (b) non-dimensional fluke-beat amplitude as functions of swimming speed per body length for several odontocete cetaceans. Adapted from \citealt{rohr2004}, and reproduced from \citealt{Floryan2016} with permission.}
\label{fig:bioData}
\end{figure}

Equations~\ref{pitch_scale} and \ref{heave_scale} indicate that the mean thrust produced by an unsteady foil is given by $\overline{T}\sim \rho s c V^2$ for both pitching and heaving motions, where $V=fa_0$.  We will assume this result may be used to estimate the thrust produced by the caudal fin.  Then,  for a constant drag coefficient, the mean body drag should scale as $\overline{D}\sim \rho s c \overline{U}_\infty^2$.   Once the animal reaches a constant swimming speed, the propulsor thrust balances the body drag, so that $ \overline{U}_\infty \sim V=fa_0$. This was partly alluded to by \citealt{bainbridge1958}, who reported that the swimming velocity of dace, trout, and goldfish obeyed a relationship $\overline{U}_\infty=\frac{1}{4}L(3f-4)$ where $L$ is the body length, although the tail amplitude as a fraction of body length was assumed constant, thus ignoring the impact of $a_0$.  Bainbridge's result then becomes $\overline{U}_\infty \sim f$ at sufficiently high tail-beat frequencies.  This observation of a linear dependence between speed and tail beat frequency, with a tail beat amplitude that remains a constant fraction of the body length, was also made by \citealt{rohr2004} for odontocete cetacean swimmers such as dolphins, porpoises, and toothed whales (see figure \ref{fig:bioData}).  Similar trends have also been seen in species of trout \citep{webb1971, webb1984} and tuna \citep{dewar1994}.  \citealt{triantafyllou1993} and \citealt{taylor2003} find that swimmers and fliers both tend towards a constant range of Strouhal number, $0.2\leq St \leq 0.4$, which also implies $\overline{U}_\infty\sim fa_0$.

As noted earlier, \citealt{gazzola2014} arrived at the same relation for thrust as we did considering only added mass effects. Equating thrust to drag, as we have here, they were able to explain biological observations over a large range of species. By also considering a Blasius drag law ($\overline{D} \sim \overline{U}_\infty ^{\,3/2}$) they were able extend their scaling to lower Reynolds number swimmers. It should also be noted that \citealt{Floryan2016} suggest that thunniform/carangiform swimmers adjust their frequency to change speed, not amplitude, because this allows them to directly manipulate their propulsive efficiency.  

\section{Conclusions}\label{S4}

By experiment and analysis, we have shown that there is little or no difference between the performances (forces and energetics) of tethered and free-swimming simple propulsors.  The analysis uses simplified scaling arguments for such propulsors in heave, pitch, and combined heave and pitch, derived from the more complete analysis given by \citealt{Floryan2016}. The thrust in heave is derived from lift-based forces, while thrust in pitch is derived from added mass forces.  At this level of modeling, both forces are governed by the lateral velocity scale $V=fa_0$, rather than the flow velocity $U_\infty$.  The experiments confirm this expectation, which for the mean velocity was anticipated by \citealt{garrick1936} for airfoils, observed in fish by \citealt{bainbridge1963}, and, more recently, explained by \citealt{gazzola2014} using only added mass forces.  We also show that this conclusion holds for streamwise oscillations in velocity, at levels far higher (up to 38\%) than considered in the past, and it extends unchanged to intermittent (or burst-coast) swimming.  Our scaling approach provides an explanation for these observations.  We further show that the structure of the wake is related to that seen in startup flows, and advance a simple physical explanation for the nature of the wake with variations in streamwise velocity, in terms of structure and mean momentum distribution. The wake is unchanged both qualitatively and quantitatively.

Although our parameter space is limited, our observations suggest that the results of constant velocity studies can be used to make robust conclusions about swimming performance without the need to explore the free-swimming condition. We believe that this message is important for the community, where this conclusion is not widely shared. Biological measurements of thunniform swimmers appear to support this conclusion, and the observations by \citealt{gazzola2014} suggest that it may extend even beyond non-thunniform swimmers. Further studies may identify the importance of the free-swimming condition in schooling/rearrangement, or escape and predation scenarios.

This work was supported by ONR Grant N00014-14-1-0533 (Program Manager Robert Brizzolara).

\section*{Appendix A}

Consider the dynamics of a freely swimming fish where the drag from the body and the thrust from the propulsor are distinct. The periodic forward swimming speed is given by
\begin{equation}
  m\frac{du}{dt} = T - D,
\end{equation}
where $m$ is the mass of the fish, $u$ is its forward swimming speed, $T$ is the thrust produced by its propulsor (for example, its caudal fin), and $D$ is the fluid drag experienced by the fish. For illustrative purposes, let $D = \frac{1}{2}\rho u^2 A_w C_D$, where $\rho$ is the density of the fluid, $A_w$ is the wetted area of the fish, and $C_D$ is a constant drag coefficient. Once the system settles on its periodic orbit, the speed and thrust may be written in terms of a Fourier series as
\begin{equation}
  u = \sum_{n = -\infty}^{\infty}u_n e^{i \omega nt}, \qquad T = \sum_{n = -\infty}^{\infty}T_n e^{i\omega nt},
\end{equation}
where $u_{-n} = u_n^*$ and $T_{-n} = T_n^*$ are the reality conditions with $^*$ denoting complex conjugation, and $\omega$ is the angular frequency of motion. The modal equations are 
\begin{align}
  n &= 0: \qquad u_0^2 = \frac{2T_0}{\rho A_w C_D} - \sum_{n \neq 0} u_n u_n^*, \\
  n &\neq 0: \qquad m \hspace{.12em} i\omega \hspace{.12em} nu_n = T_n - \frac{1}{2}\rho A_w C_D \sum_{k = -\infty}^{\infty} u_ku_{n - k}.
\end{align}
We see that the mean speed is given by the balance of mean thrust and mean drag, minus a modification due to the higher harmonics. This modification is in general small because the higher harmonics decay approximately as $1/n$ and only the even modes of thrust are nonzero for symmetric motions of the propulsor.

\section*{Appendix B}
Here we show that inertia of the foil has no impact on the mean forces or power. Consider a foil of mass $m$ held at its leading edge. The leading edge is located at position $(x,y)$, its center of mass is located a distance $d$ from the leading edge, and its moment of inertia about the leading edge is $I$. We move the foil in some periodic fashion in $(x,y,\theta)$. The position of the center of mass is $(x + d\cos\theta, y + d\sin\theta)$, its velocity is $(\dot{x} - d\, \dot{\theta}\sin\theta, \dot{y} + d \, \dot{\theta}\cos\theta)$, and its acceleration is $(\ddot{x} - d \, \ddot{\theta}\sin\theta - d \, \dot{\theta}^2\cos\theta, \ddot{y} + d \, \ddot{\theta}\cos\theta - d \, \dot{\theta}^2\sin\theta)$. 

The forces and moment due to the inertia of the foil are then given by
\begin{align*}
  \mathbf{F} &= m\mathbf{a}_{cm} = m(\ddot{x} - d\ddot{\theta}\sin\theta - d\dot{\theta}^2\cos\theta, \ddot{y} + d\ddot{\theta}\cos\theta - d\dot{\theta}^2\sin\theta), \\
  M_{le} &= \mathbf{r}_{cm/le} \times m\mathbf{a}_{cm} + I\ddot{\theta} \\
  &= m\left[d\cos\theta\left(\ddot{y} + d\ddot{\theta}\cos\theta - d\dot{\theta}^2\sin\theta\right) - d\sin\theta\left(\ddot{x} - d\ddot{\theta}\sin\theta - d\dot{\theta}^2\cos\theta\right)\right] + I\ddot{\theta}.
\end{align*}

The mean force due to inertia is found by integrating over a full period $T$, so that
\begin{align*}
\overline{ \mathbf{F} } &= \frac{1}{T}\int_0^T \mathbf{F}\, dt \\
  &= \frac{1}{T}\int_0^T m\mathbf{a}\, dt = \frac{m}{T}\int_0^T \frac{d\mathbf{v}}{dt}\, dt = \frac{m}{T}\int_{\mathbf{v}(0)}^{\mathbf{v}(T)} d\mathbf{v}' = \frac{m}{T}\left[\mathbf{v}(T) - \mathbf{v}(0)\right] = 0,
\end{align*}
due to periodicity. The inertia of the propulsor therefore does not affect our measurements of mean forces. 

The mean power due to inertia is given by
\begin{align*}
\overline{ P } &= \frac{1}{T}\int_0^T P\, dt  \  = \  \frac{1}{T}\int_0^T \left(M\dot{\theta} + F_y\dot{y}\right)\, dt \\
  &= \frac{1}{T}\int_0^T \Big\{m\dot{\theta}\left[d\cos\theta\left(\ddot{y} + d\ddot{\theta}\cos\theta - d\dot{\theta}^2\sin\theta\right) - d\sin\theta\left(\ddot{x} - d\ddot{\theta}\sin\theta - d\dot{\theta}^2\cos\theta\right)\right] \\
  & + I\ddot{\theta}\dot{\theta} + m\dot{y}\left(\ddot{y} + d\ddot{\theta}\cos\theta - d\dot{\theta}^2\sin\theta\right)\Big\}\, dt \\
  &= \frac{1}{T}\int_0^T\left[\frac{1}{2}I\frac{d}{dt}(\dot{\theta}^2) + \frac{1}{2}m\frac{d}{dt}(\dot{y}^2) + \frac{1}{2}md^2\frac{d}{dt}(\dot{\theta}^2) + md\frac{d}{dt}(\dot{y}\dot{\theta}\cos\theta) - md\ddot{x}\dot{\theta}\sin\theta\right]\, dt \\
  &= -\frac{md}{T}\int_0^T \ddot{x}\dot{\theta}\sin\theta\, dt.
\end{align*}
We note that the mean power is nonzero only if there are simultaneous streamwise and pitching motions. 

For the motions where the streamwise position of the propulsor changes with twice the frequency of the actuation, the integrand is an odd periodic function, and so the integral is exactly zero. For the burst-coast motions, the integrand is nonzero only during the bursting phase. Re-centering about the middle of the bursting phase of the motion, the integrand is an odd function, and so the integral is exactly zero. Thus for the motions considered in this work, the inertia of the propulsor does not affect our measurements of mean power. 

\bibliographystyle{jfm}

\end{document}